\documentclass[10pt,twocolumn,floatfix,altaffilletter,superscriptaddress,tightenlines,showpacs,showkeys,preprintnumbers,nofootinbib]{revtex4-1}
\pdfoutput=1
\usepackage{hyperref}
\usepackage{amsmath,amssymb}
\usepackage{epsfig} 
\usepackage{graphicx}        
\usepackage{url}
\usepackage{color}
\usepackage{multirow}
\usepackage{placeins}
\usepackage[dvipsnames]{xcolor}
\usepackage{braket}
\usepackage{float}
\usepackage{slashed}

\hypersetup{colorlinks,linkcolor={blue},citecolor={teal},urlcolor={violet}}
 
\allowdisplaybreaks

\bibpunct{[}{]}{,}{n}{}{,}

\def\beq{\begin{equation}}
\def\eeq{\end{equation}}
\def\bea{\begin{eqnarray}}
\def\eea{\end{eqnarray}}
\makeatletter

\newcommand{\INFN}{INFN - Sezione di Napoli, Complesso Universitario Monte S. Angelo, I-80126 Napoli, Italy}
\newcommand{\UNINA}{Dipartimento di Fisica ``Ettore Pancini'', Università degli studi di Napoli ``Federico II'',\\ Complesso Universitario Monte S. Angelo, I-80126 Napoli, Italy}
\newcommand{\SSM}{Scuola Superiore Meridionale, Università degli studi di Napoli ``Federico II'', Largo San Marcellino 10, I-80138 Napoli, Italy}
\newcommand{\NAN}{Department of Physics and Institute of Theoretical Physics, Nanjing Normal University, Nanjing, 210023, China}

\newcommand{\documenttitle}{Probing flavored regimes of leptogenesis with gravitational waves from cosmic strings}

\begin{document}

\title{\documenttitle}
\author{Marco Chianese}
    \email{marco.chianese@unina.it}
    \affiliation{\UNINA}
    \affiliation{\INFN}
\author{Satyabrata Datta}
    \email{amisatyabrata703@gmail.com}
    \affiliation{\NAN}
\author{Gennaro Miele}
    \email{miele@na.infn.it}
    \affiliation{\UNINA}
    \affiliation{\INFN}
    \affiliation{\SSM}
\author{Rome Samanta}
    \email{samanta@na.infn.it}
    \affiliation{\SSM}
    \affiliation{\INFN}
\author{Ninetta Saviano}
    \email{nsaviano@na.infn.it}
    \affiliation{\INFN}
    \affiliation{\SSM}

\begin{abstract} 
Cosmic strings radiate detectable gravitational waves in models featuring high-scale symmetry breaking, e.g., high-scale leptogenesis.  In this Letter, for the first time, we show that different flavored regimes of high-scale leptogenesis can be tested with the spectral features in cosmic string-radiated gravitational waves. This is possible if the scalar field that makes right-handed neutrinos massive is feebly coupled to the Standard Model Higgs. Each flavored regime, sensitive to low-energy neutrino experiments, leaves a marked imprint on the gravitational waves spectrum. A three-flavor and a two-flavor regime could be probed by a characteristic fall-off of the gravitational wave spectrum at the LISA-DECIGO-ET frequency bands with preceding scale-invariant amplitudes bounded from above and below. We present Gravitational Waves windows for Flavored Regimes of Leptogenesis (GWFRL) testable in the upcoming experiments. We also provide the first construction of a leptogenesis framework where a testable distinction of flavor regimes is possible without constraining the flavor structure of the theory.
\end{abstract}

\maketitle
{\bf Introduction:} Within the framework of the seesaw mechanism of light neutrino masses, leptogenesis \cite{Fukugita:1986hr} is a simple process for obtaining the observed baryon asymmetry of the universe \cite{Planck:2015fie}. In this process,  CP asymmetric and out-of-equilibrium decays of heavy right-handed neutrinos (RHNs) to lepton and the Standard Model (SM) Higgs doublets, create a lepton asymmetry, which is then processed to baryon asymmetry by sphalerons \cite{Kuzmin:1985mm}. In the thermal leptogenesis scenario, the lepton asymmetry is produced at a temperature $T_{\rm lepto}\sim M_ N$, where $M_N$ is the mass of the decaying RHN -- also referred to as the scale of thermal leptogenesis~\cite{Davidson:2008bu,Buchmuller:2004nz}. Despite being a simple mechanism, leptogenesis operating on scales $M_N> \mathcal{O}\left(\rm TeV\right)$ is not directly testable, e.g., with collider experiments. Accurate incorporation of flavor effects \cite{Abada:2006ea,Nardi:2006fx,Blanchet:2006be} nonetheless opens up the opportunity to test high-scale leptogenesis with low-energy neutrino observables, specifically with the leptonic CP violation \cite{Pascoli:2006ci} -- a measurement of which is one of the primary goals of the ongoing neutrino experiments such as T2K \cite{T2K:2018rhz} and NO$\nu$A \cite{NOvA:2018gge}. Depending on the interaction strengths of the charged lepton flavors, there exist three distinct regimes of high-scale leptogenesis:  $M_N\gtrsim 10^{12}~{\rm GeV}$ (one-flavor/vanilla, 1FL), $10^9~{\rm GeV}\lesssim M_N\lesssim 10^{12}~{\rm GeV}$ (two-flavor, 2FL), and  $M_N\lesssim 10^{9}~{\rm GeV}$ (three-flavor, 3FL). While the 1FL regime is not sensitive to neutrino mixing parameters, the 2FL and 3FL regimes can offer low-energy neutrino phenomenology, which however may not be unequivocal, unless the flavor structure of the theory is constrained by invoking symmetries (see, e.g., Ref.s~\cite{Akhmedov:2003dg,Bertuzzo:2009im}).

Lately, there have been efforts to obtain a direct possible signature of high-scale leptogenesis with primordial stochastic Gravitational Waves background (GWs) by focusing on the RHN sector. This is promising because, unlike electromagnetic radiation, GWs traverse the early universe unimpeded with unerring information about their origin. Therefore, no matter how high the leptogenesis scale is, it can be probed with GWs. The idea arises from contemplating the origin of RHN mass at a more fundamental level~\cite{Dror:2019syi,Blasi:2020wpy,Samanta:2020cdk,Datta:2020bht}. Indeed, in Ref.~\cite{Dror:2019syi} it has been argued that the RHNs become massive because of the spontaneous breaking of a $U(1)_{B-L}$ gauge invariance, which many Grand Unified Theories (GUT) might embed \cite{Davidson:1978pm,Marshak:1979fm,Buchmuller:2013lra,Buchmuller:2019gfy}. Besides generating RHN mass, a gauged $U(1)_{B-L}$ breaking inevitably produces cosmic strings (here we consider only stable cosmic strings~\cite{Buchmuller:2023aus}) that radiate gravitational waves, which may be a signature of leptogenesis. Let us consider the $U(1)_{B-L}$ invariant RHN mass term $f_N \overline{N_R} \Phi N^C_R$, where $f_N$ is the Yukawa coupling, $N_R$ is the RHN field and $\Phi$ being the $B-L$ scalar with vacuum expectation value $v_\Phi$. After the $U(1)_{B-L}$ breaking, the RHNs obtain mass given by $M_N=f_N v_\Phi$, cosmic strings are formed~\cite{Kibble:1976sj,Hindmarsh:1994re,Jeannerot:2003qv} and radiate GWs with a scale-invariant spectrum at higher frequencies with an overall amplitude $\Omega_{\rm GW}\propto v_\Phi$~\cite{Vilenkin:1981bx,Vachaspati:1984gt}. While the scenario is interesting, note however, that the GW spectrum being dependent only on $v_\Phi$ is not so sensitive to $M_N$ because of the additional free parameter $f_N$. As such, a GW spectrum for $v_\Phi=10^{14}$ GeV may correspond to  3FL for $f_N=10^{-6}$ as well as 2FL for $f_N=10^{-4}$. Therefore, a more realistic GW signature of leptogenesis requires dependence on both parameters $v_\Phi$ and $f_N$, and hence also on $M_N$. 

In this Letter, we show that the $M_N$-dependent GW spectrum and a concrete link between GWs and high-scale leptogenesis can be achieved in the $U(1)_{B-L}$ model with a peculiar scalar field dynamics~\cite{Chianese:2024nyw}. After the phase transition, $\Phi$ rolls down to its true vacuum and oscillates coherently to provide an early matter-domination when it is long-lived. The lifetime of the scalar field is determined by $v_\Phi$ and $f_N$, and therefore,  $M_N$ controls the beginning and the end of the matter-domination. Differently from Ref.~\cite{Dror:2019syi}, the amplitude and the spectral features of the GWs radiated from cosmic strings are now imprinted by the $M_N$-dependent matter domination, thus carrying information on $M_N$, and therefore, on different regimes of flavored leptogenesis. In particular, the scale invariance in the GW spectrum at higher frequencies breaks at a frequency $f_{\rm dec}$ which is determined by the end of the matter domination at the temperature $T_{\rm dec}(M_N)$ (see Fig.~\ref{fig:GW}). We point out that such a scenario allows us to probe the RHN mass scale markedly, in agreement with all the relevant constraints on the parameter space (see Fig.~\ref{fig:GWFRL}). In particular, the prominent ones are: {\it i)} the bound from the recent observations of nHz GWs by Pulsar Timing Arrays (PTA)~\cite{NANOGrav:2023gor,EPTA:2023fyk,Reardon:2023gzh,Xu:2023wog,EPTA:2023xxk}  which disfavor GWs from stable cosmic strings~\cite{NANOGrav:2023hvm}; {\it ii)} the condition that the cosmic strings radiate predominantly GWs than particles~\cite{Matsunami:2019fss,Auclair:2019jip}; {\it iii)} the requirement that the $\Phi$ particles decay before the Big-Bang-Nucleosynthesis (BBN)~\cite{Cyburt:2015mya,Hasegawa:2019jsa}; and {\it iv)} the scalar field $\Phi$ is long-lived. We highlight that if the break of the GW spectrum occurs in the LISA (DECIGO) frequency band, it is a signature (most likely signature) of a 3FL regime (see Fig.~\ref{fig:MN}). On the other hand, a break of scale invariance in ET would correspond to both regimes,  but with less signal strength for a 3FL. Remarkably, the amplitudes and the spectral break frequencies of the GW spectra predicted by our model are very well delimited and they are all within reach of planned gravitational wave experiments such as LISA, DECIGO, and ET~\cite{lisa,decigo,et}. Our results not only ensure  testable GW signatures of different flavor regimes but also imply that the latter can be probed without constraining the flavor space of the leptogenesis Lagrangian.

{\bf The $M_N$-dependent matter domination and constraints:} The seesaw Lagrangian for neutrino masses and leptogenesis is given by
\bea
-\Delta \mathcal{L}\subset f_{D i\alpha} \overline{N_{Ri} }\tilde{H}^\dagger L_{\alpha}+\frac{1}{2}f_{Ni}\overline{N_{Ri}}\Phi N_{Ri}^C+{\rm h.c.}\,,\label{seesawlag}
\eea
where $f_D$ is the neutrino Dirac Yukawa coupling and $H(L)$ is the SM Higgs (lepton doublet). The fields $N_R$ and $L_\alpha$ have $B-L$ charge $-1$, whereas the same charges for $\Phi$ and $H$ are $2$ and $0$, respectively. The scalar field dynamics is dictated by the finite temperature potential restoring the symmetry at higher temperatures given by \cite{Linde:1978px,Kibble:1980mv}
\bea
V(\Phi,T)=\frac{\lambda}{4}\Phi^4+D(T^2-T_0^2)\Phi^2-ET\Phi^3 \,,\label{tmpdp}
\eea
where 
$D,E$ and $T_0$ are functions of $B-L$ gauge coupling $g^\prime$, the self-interaction coupling $\lambda$ and the vacuum expectation value $v_\Phi=m/\sqrt{\lambda}$, which is determined by the zero temperature potential $V(\Phi,0)=-\frac{m^2}{2}\Phi^2+\frac{\lambda}{4}\Phi^4$. The last term in Eq.~\eqref{tmpdp} generates a potential barrier causing a secondary minimum at $\Phi\neq0$, which becomes degenerate with the $\Phi=0$ one at $T=T_c$. The potential barrier vanishes at $T_0 ~(\lesssim T_c)$, and the minimum at $\Phi=0$ becomes a maximum. The transition from $\Phi = 0$ to $\Phi = v_\Phi$ can be treated as a second-order transition with an extremely quick disappearance of the potential barrier if the order parameter $\Phi_c/T_c\ll 1$. Such a smooth transition can be obtained approximately within the ballpark of $\lambda \simeq{ g^\prime}^3$ and for ${ g^\prime} \lesssim 10^{-2}$ \cite{Blasi:2020wpy,Datta:2023vbs}.  Once $\Phi$ rolls down, it oscillates around $v_{\Phi}$ behaving as matter~\cite{Masso:2005zg,Datta:2022tab,Chianese:2024nyw}. Three important observations are in order. First, the $U(1)_{B-L}$ breaking scale can be as high as the GUT scale $\Lambda_{\rm GUT}$, implying $v_\Phi \lesssim 5 \times 10^{15}~{\rm GeV}$. Second, the Hubble friction at $T_c$ should also be negligible for the rolling of the field down to $v_\Phi$; this can be achieved by considering $m_\Phi$ larger than the Hubble parameter $\mathcal{H}(T_c)$. Third, RHNs become massive after the phase transition at $T=T_c$. Therefore, the scale of leptogenesis is bounded from above as $T_{\rm lepto} \sim M_N\lesssim T_c\simeq  2\sqrt{g^\prime} v_\Phi$. 

We aim to make $\Phi$ long-lived to induce a matter-domination epoch with a lifetime determined by $v_\Phi$ and $f_N$. In high-scale leptogenesis scenario, the coupling $f_N$ should be generically large to have $M_N$ comparable with $v_\Phi$. In such a case, if $m_\Phi > 2 M_N$, tree-level decay of $\Phi$ to RHN pairs ($\Phi\rightarrow N N$) would be too quick to make $\Phi$ long-lived. This restricts ourselves to the case $M_N>m_\Phi=\sqrt{2 \lambda}v_\Phi$. Therefore, along with $ M_N\lesssim T_c$, the $\Phi$ scalar is long-lived within the following range
\bea
m_\Phi < M_N \lesssim T_c~~\implies~~\sqrt{2g^\prime}g^\prime  <f_N\lesssim 2\sqrt{g^\prime} \,, \label{rhmass_b}
\eea
where we have assumed $\lambda={g^\prime}^3$. In this case, the mass of $\Phi$ is suppressed by a factor $\sqrt{g^\prime}$ compared to the $M_{Z^\prime}=\sqrt{2}g^\prime v_\Phi$, thus kinematically forbidding the tree-level decays of $\Phi$ to $Z^\prime$ pairs. 
Note interestingly that, even if defined to vanish at a very high scale, at lower energies, the seesaw Lagrangian in Eq.~\eqref{seesawlag} can provide a nonvanishing effective coupling $\lambda_{\Phi H, {\rm 1-loop}} \sim f_D^2 f_N^2 /(2\pi^2)$ at one-loop to trigger $\Phi\rightarrow HH $ decay ~\cite{Gross:2015bea,Enqvist:2016mqj,Chianese:2024nyw}. Therefore, despite $f_N$ being large, we can have a lower decay rate proportional to $\lambda_{\Phi H, \text{1-loop}}^2$. This decay involves both the RHN mass and the active neutrino mass $m_\nu$ via the Dirac Yukawa coupling $f_D=\sqrt{m_\nu f_N v_\Phi/v_H^2}$. In our discussion, we consider this radiative $\Phi\rightarrow HH $ decay with the assumption that $\lambda_{\Phi H, \text{1-loop}}\gg \lambda_{\Phi H, \text{tree}}$ (This the crucial theoretical identification in the $U(1)_{B-L}$ model, that makes GW spectrum sensitive to $M_N$). The decay rate is given by~\cite{Gross:2015bea,Enqvist:2016mqj,Chianese:2024nyw}
\bea
\Gamma_\Phi^{HH} \simeq \Gamma_0\,\frac{f_N^6}{\lambda}\left(\frac{v_\Phi}{\rm 10^{13}~GeV}\right)^2\left(\frac{m_\Phi}{\rm 10^8~GeV}\right)\ln^2\left(\frac{\Lambda}{\tilde{\Lambda}}\right) \,, \label{gamma_dec}
\eea
where the running of the one-loop effective coupling is determined by the UV cut-off scale, e.g., $\Lambda\simeq \Lambda_{\rm GUT}$ and the relevant physical scale $\tilde{\Lambda}$. We have $\Gamma_0=1.3 \times 10^{-2}~{\rm GeV}$ assuming $m_\nu\simeq 0.01$~eV and the SM Higgs vacuum expectation value $v_H=174$~GeV.

The evolution of the energy density of $\Phi$ and the duration of the matter-dominated epoch, which ends when $\Phi$ decays~\cite{Chianese:2024nyw}, can be tracked by solving the Friedmann equations (see Supplemental Material~\ref{SM:evo} for further details). We can compute the decay temperature of the scalar field by recursively solving the implicit equation $\Gamma_\Phi^{HH} (T_{\rm dec}) \simeq \mathcal{H}(T_{\rm dec})$ taking $\tilde{\Lambda} \simeq T_{\rm dec}$ in Eq.~\eqref{gamma_dec}. The scalar must decay before BBN, i.e., $T_{\rm dec}\gtrsim T_{\rm BBN}\sim 10 ~ {\rm MeV}$. Moreover, because the scalar field is long-lived, it produces entropy which dilutes any pre-existing relic. Therefore, in the computation of baryon asymmetry, the entropy dilution must be considered, because in this framework, the produced entropy can be as large as $10^6$ (see Supplemental Material \ref{SM:evo}). Finally, we mention that we can safely neglect the additional one-loop decay $\Phi\rightarrow f\bar{f}V$ (the two-body decay $\Phi \rightarrow f\bar{f}$ is suppressed due to chirality flip~\cite{Han:2017yhy}), where $f$ and $V$ are SM fermions and vector bosons. Indeed, this decay dominates over the $\Phi\rightarrow HH$ only for $g^\prime\gtrsim 10^{-2}$ and smaller values of $v_\Phi$, which are outside our parameter space.

{\bf  Imprints of flavored leptogenesis on GWs from cosmic strings:} Gravitational waves are radiated from cosmic string loops chopped off from the long strings resulting from the spontaneous breaking of the gauged $U(1)_{B-L}$ \cite{Vilenkin:1981bx,Vachaspati:1984gt}. Long strings are described by a correlation length $L=\sqrt{\mu/\rho_\infty}$, where $\rho_\infty$ is the long string energy density and $\mu $ is the string tension defined as~\cite{Hill:1987qx} 
\bea
\mu=\pi v_\Phi^2~h\left(\lambda,g^\prime\right)~~{\rm with}~~h\simeq \frac{1}{\ln({2 {g^\prime}^2}/{\lambda})} \,,
\eea
for $\lambda\ll 2 {g^\prime}^2$. In the present analysis, we take $\lambda={g^{\prime}}^3$ therefore $h\simeq 1/\ln(2 / g^\prime)$. The time evolution of a radiating loop of initial size $l_i=\alpha t_i$ is given by $l(t)=l_i-\Gamma G\mu(t-t_i)$, where $\Gamma\simeq 50$~\cite{Vilenkin:1981bx,Vachaspati:1984gt}, $\alpha\simeq 0.1$~\cite{Blanco-Pillado:2013qja,Blanco-Pillado:2017oxo}, $G=M_{\rm Pl}^{-2}$, and $t_i$ being the initial time of loop creation. The total energy loss from a loop is decomposed into a set of normal-mode oscillations with frequencies $f_k=2k/l_k=a(t_0)/a(t)f$, where $k=1,2,3...k_{\rm max}$, $f$ is the present-day frequency at $t_0$, and $a$ is the scale factor. The total GW energy density is computed by summing all the $k$ modes giving~\cite{Blanco-Pillado:2013qja,Blanco-Pillado:2017oxo} 
\bea
\Omega_{\rm GW}= \sum_{k=1}^{k_{\rm max}}\frac{2k \mathcal{F}_\alpha G\mu^2 \Gamma_k}{f\rho_c}\int_{t_{i}}^{t_0} \left[\frac{a(t)}{a(t_0)}\right]^5 n_\omega\left(t,l_k\right){\rm d}t \,,\label{gwcs1}
\eea
where $\rho_c$ is the critical energy density of the universe, $\mathcal{F}_\alpha\simeq 0.1$ is an efficiency factor~\cite{Blanco-Pillado:2013qja} and $n_\omega\left(t,l_k\right)$ is the loop number density. The latter can be computed from the velocity-dependent-one-scale model as~\cite{Martins:1996jp,Martins:2000cs,Sousa:2013aaa,Auclair:2019wcv}
\bea
n_\omega(t,l_{k})=\frac{A_\beta}{\alpha}\frac{(\alpha+\Gamma G \mu)^{3(1-\beta)}}{\left[l_k(t)+\Gamma G \mu t\right]^{4-3\beta}t^{3\beta}} \,, \label{genn0}
\eea
where $\beta=2/3(1+\omega)$ with $\omega$ being the equation of state parameter of the universe, and $A_\beta =5.4$ ($A_\beta = 0.39$) for radiation-dominated (matter-dominated) universe~\cite{Auclair:2019wcv}. The quantity $\Gamma_k={\Gamma k^{-\delta}}/{\zeta(\delta)}$ quantifies the emitted power in $k$-th mode, with $\delta=4/3$ ($\delta=5/3$) for loops containing cusps (kinks) \cite{Damour:2001bk}.
\begin{figure}[t!]
    \centering
    \includegraphics[width=\columnwidth]{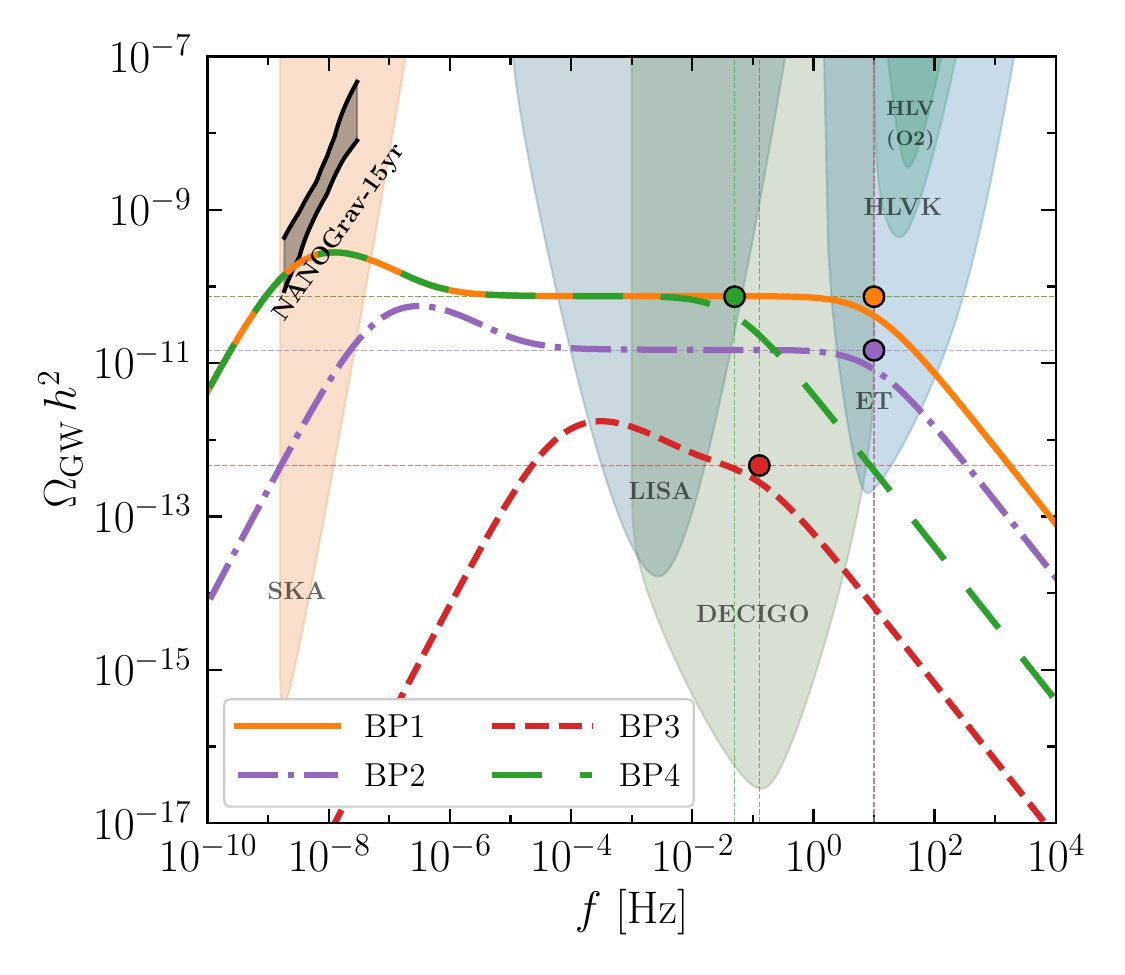}
    \caption{The GW spectrum from the cosmic strings predicted by our model for four benchmark cases with $g^\prime=10^{-4}$: BP1 (2FL), $M_N=1.6 \times 10^9~{\rm GeV}$ and $f_N=1.4 \times 10^{-5}$; BP2 (3FL), $4.2 \times 10^8~{\rm GeV}$ and $f_N=1.8 \times 10^{-5}$; BP3 (3FL), $5.2 \times 10^6~{\rm GeV}$ and $f_N=6.8 \times 10^{-6}$; BP4 (3FL), $2.7 \times 10^8~{\rm GeV}$ and $f_N=2.2 \times 10^{-6}$. The points show the calculations of the GW plateau amplitude $\Omega_{\rm GW}^{\rm plt}$ and the break frequency $f_{\rm dec}$ according to Eq.s~\eqref{flp1} and~\eqref{brk}, respectively. BP1 and BP2 correspond to equal $\Omega_{\rm GW}^{\rm plt}$ but different $f_{\rm dec}$, while BP1 and BP4 to equal $f_{\rm dec}$ but different $\Omega_{\rm GW}^{\rm plt}$. BP3 represents the weakest allowed signal for $g^\prime = 10^{-4}$ (see Fig.~\ref{fig:GWFRL}).}
    \label{fig:GW}
\end{figure}
The integral in Eq.~\eqref{gwcs1} is subjected to two Heaviside functions $\Theta(t_i-t_{\rm fric})\Theta(t_i-l_{c}/\alpha)$, each setting a  cut-off at a very high frequency $f_*$, beyond which the GW spectrum falls as $f^{-1}$. A similar cut-off owing to inflation-diluted strings~\cite{Guedes:2018afo} is not so relevant to our context because we consider thermal leptogenesis. The quantity $t_{\rm fric}$ represents the time until which the motion of the string network is damped by friction~\cite{Vilenkin:1991zk}, and  $l_c$ is a critical length above which GW emission is dominant over particle production as shown by high-resolution numerical simulations~\cite{Matsunami:2019fss,Auclair:2019jip} (see also Ref.~\cite{Baeza-Ballesteros:2024otj}). The critical length is $l_c\simeq\delta_w (\Gamma G \mu)^{-\gamma}$, where $\delta_w=(\sqrt{\lambda}v_\Phi)^{-1}$ is the width of the string, and $\gamma=2$ ($\gamma=1$) for loops containing cusps (kinks). We shall neglect the cut-offs due to the kinks and friction as they are weaker than those from cusps. Here we deal with small values of $\lambda$ leading to ``thick'' strings, which are not conventional while studying GWs signatures from cosmic strings in BSM models~\cite{Dror:2019syi}. In this case, the particle production cut-off in the GW spectrum is expected to be much stronger than the one usually considered for ``thin'' strings ($\lambda\simeq 1$). Thus, our scenario is also the first concrete BSM model that discusses the results of Ref.s~\cite{Matsunami:2019fss,Auclair:2019jip} for small $\lambda$. Note however that, we generalized the simulation results obtained for large couplings ($\lambda,g^\prime$) by replacing $\delta_w=(\sqrt{\lambda}v_\Phi)^{-1}$ in $l_c$. But in principle, Eq.(11) of Ref.~\cite{Matsunami:2019fss} would exhibit coupling dependence which is not trivial to deduce without numerical simulation. Therefore, the particle production cut-off discussed above should be considered conservative. Note however that the strings predicted in this model effectively evolve as thin strings because the condition $\delta_w \ll \mathcal{H}^{-1}$ is always satisfied (see Supplemental Material \ref{SM:strings}). In the following, we present the results for the dominant $k=1$ mode which captures the qualitative features of our analysis.
\begin{figure}[t!]
    \centering
    \includegraphics[width=\columnwidth]{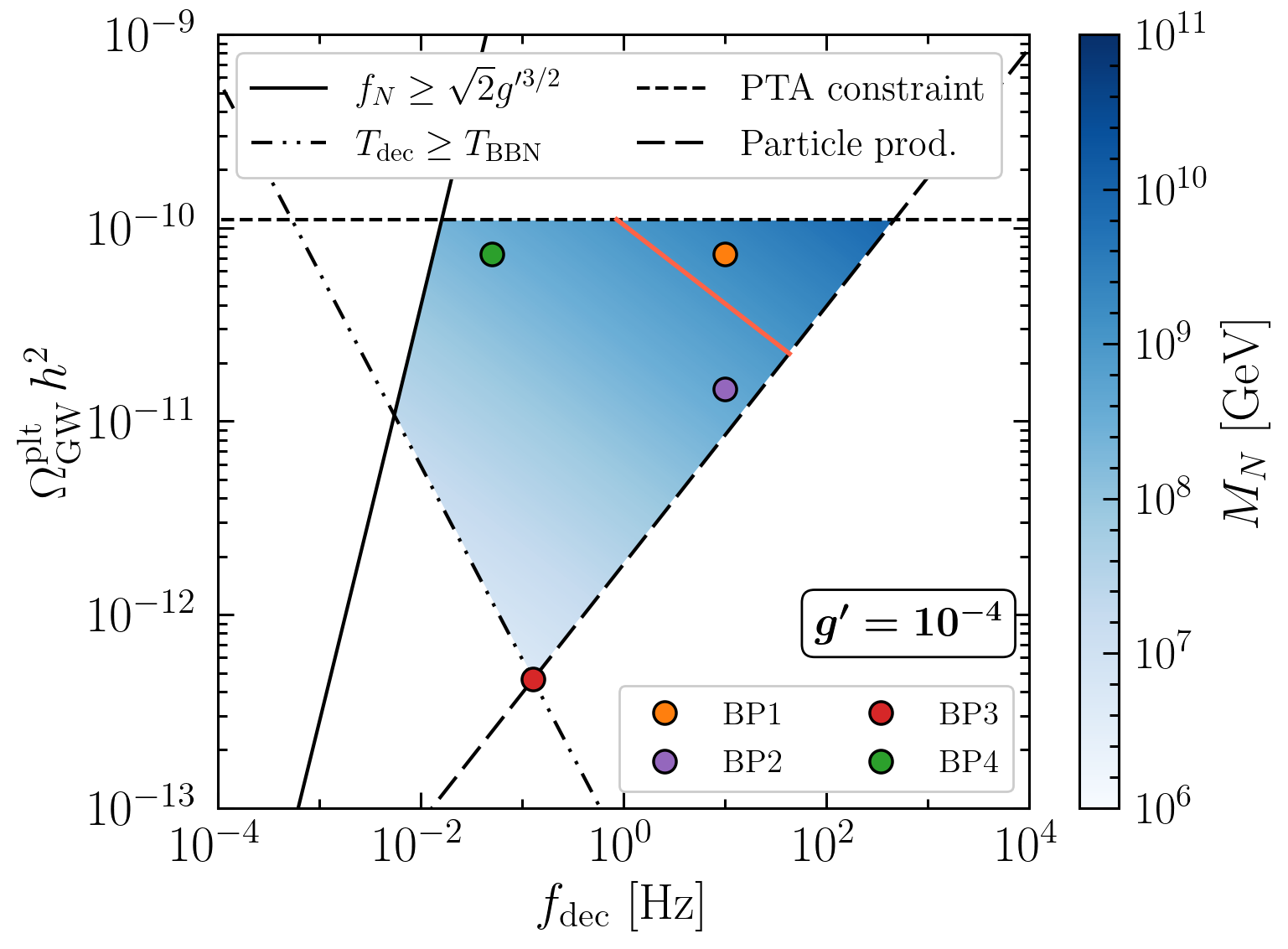}
    \caption{The allowed parameter space (GWFRL window) in the $f_{\rm dec}$-$\Omega_{\rm GW}^{\rm plt}\,h^2$ plane for $g^\prime = 10^{-4}$ with the leptogenesis scale $M_N$ color-coded. The different black lines display the constraints according to the discussed model requirements and the PTA measurements. The different points correspond to the benchmark cases displayed in Fig.~\ref{fig:GW}. The orange solid line marks the transition between three-flavor and two-flavor leptogenesis regimes.}
    \label{fig:GWFRL}
\end{figure}

In Fig.~\ref{fig:GW}, we show the GW spectrum from Eq.~\eqref{gwcs1} (for $k=1$) predicted by our model for four benchmark scenarios mentioned in the caption. Without any intermediate matter-dominated epoch, the GW spectrum has two main features: {\it i)} a low-frequency peak owing to the GW radiation from the loops that originate in the radiation epoch and decay in the standard matter epoch, and {\it ii)} an almost scale-invariant plateau at high frequencies as~\cite{Blanco-Pillado:2013qja,Blanco-Pillado:2017oxo,Sousa:2020sxs}
\bea
\Omega_{\rm GW}^{\rm plt} = \frac{128\pi\mathcal{F}_\alpha G\mu}{9\zeta(\delta)}\frac{A_R}{\epsilon_R}\Omega_R\left[(1+\epsilon_R)^{3/2}-1\right] \,, \label{flp1}
\eea
that arises from loop dynamics in the radiation-dominated epoch only. In Eq.~\eqref{flp1}, 
$\epsilon_R=\alpha/\Gamma G\mu \gg 1$, $A_R\equiv A_\beta\simeq 5.4$ and $\Omega_R\sim 9\times 10^{-5}$. Note that $\Omega_{\rm GW}^{\rm plt}\propto \sqrt{\mu}\propto v_\Phi$, implying the larger the symmetry breaking scale, the stronger the amplitude of GW. In our scenario, which features instead an intermediate matter-dominated epoch, the plateau breaks at a high frequency $f_{\rm dec}$, beyond which the spectrum falls as $\Omega_{\rm GW}(f>f_{\rm dec})\propto f^{-1}$ (see Fig.~\ref{fig:GW}). The analytical expression for this spectral break frequency $f_{\rm dec}$ can be estimated as  \cite{Cui:2018rwi}  
\bea
f_{\rm dec}\simeq 0.45\,{\rm Hz}\left(\frac{10^{-12}}{G\mu}\right)^{1/2}\left(\frac{T_{\rm dec}}{\rm GeV}\right)\,.\label{brk}
\eea
To robustly claim the spectral fall as a signature of $M_N$-dependent matter domination, we must require $t_{\rm dec} > l_c/\alpha$. This is equivalent to imposing a constraint on the parameter space such that $T_{\rm cusp}/T_{\rm dec}>1$, where $T_{\rm cusp}$ can be derived as 
\bea
T_{\rm cusp}\simeq 25\,{\rm GeV} \left(\frac{G \mu}{10^{-12}}\right)^{5/4}\left(\frac{g^\prime}{10^{-4}}\right)^{3/4}h(g^\prime)^{-1/4}\,.
\label{c_cusp}
\eea
For loops containing kinks, one can obtain a similar constraint, which however is less stringent than Eq.~\eqref{c_cusp}. The parameter space of our model is also constrained by the recent PTA observation of stochastic GWs, which disfavors stable cosmic strings due to a different spectral slope~\cite{NANOGrav:2023hvm}. Thus, PTA data require $\Omega_{\rm GW}\lesssim \Omega_{\rm GW}^{\rm PTA}$ at $f=30$ nHz-- giving $G\mu\lesssim 7.0\times 10^{-11}$.

We note that there could be an additional contribution to the GW spectrum at very high frequencies from the cosmic string loops in the first radiation epoch before $\Phi$ starts to dominate. However, the amplitude at higher frequencies is suppressed by the entropy production and the particle production cut-off.
\begin{figure}[t!]
    \centering
    \includegraphics[width=\columnwidth]{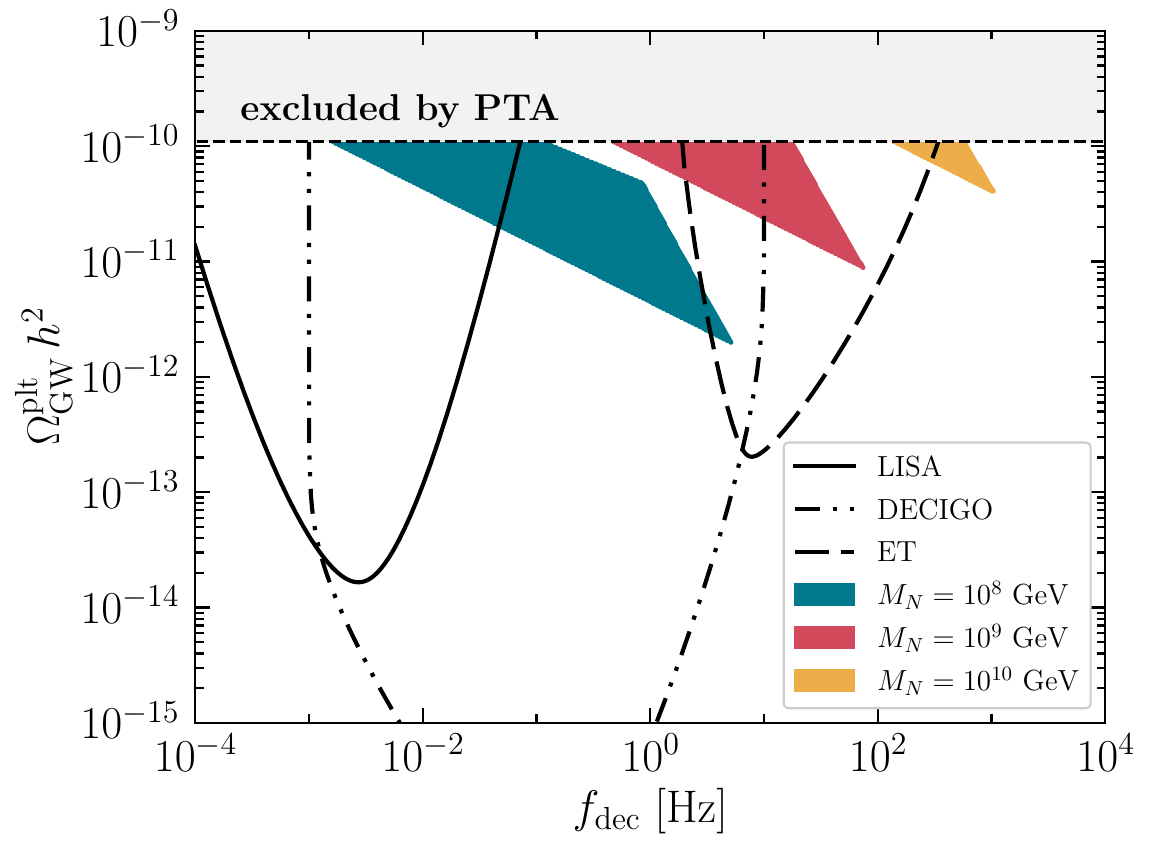}
    \caption{The GWFRL windows for different values of the leptogenesis scale $M_N$, scanning over the model parameters $g^\prime$ and $f_N$. The black lines show the sensitivity of LISA, DECIGO and ET detectors (black lines). The grey shaded region is excluded by current PTA data.}
    \label{fig:MN}
\end{figure}

{\bf  Results and discussion:} 
In Fig.~\ref{fig:GWFRL}, we show the allowed parameter space for $g^\prime = 10^{-4}$, in agreement with all the model conditions and constraints discussed above. For a given gauge coupling, the spectral break frequency $f_{\rm dec}$ and the GW plateau amplitude $\Omega_{\rm GW}^{\rm plt}$ are uniquely determined by the model parameters $M_N$ and $f_N$. Usually, they are limited from above by the PTA data. However, in our model, they are also bounded from below by the BBN constraint on the matter-dominated epoch and by the requirement that the spectral break is not due to the particle production from cosmic strings. Scanning over different values of the gauge coupling (see Supplemental Material~\ref{SM:gauge} for the gauge coupling dependence of the parameter space), we find that our scenario implies $10^{-3} \lesssim f_{\rm dec}/{\rm Hz} \lesssim 10^3$ and $10^{-13}\lesssim\Omega_{\rm GW}^{\rm plt}h^2\lesssim 10^{-10}$, which we define as the Gravitational Waves windows for Flavored Regimes of Leptogenesis (GWFRL). These ranges, within which a robust GW signature of high-scale leptogenesis is achieved, will be fully explored by planned GW experiments such as LISA, DECIGO and ET. These GWFRL windows also trivially satisfy two important constraints. First, $G\mu<6.3\times 10^{-3} g^\prime h(g^\prime)$ due to the Hubble friction being always less than $m_\Phi$ at $T\simeq T_c$. Second, the ratio of the $\Phi$ vacuum energy to the radiation at $T_c$; $\rho_\Phi(T_c)/\rho_{R}(T_c)=4.7\times 10^{-4}g^\prime$, is always smaller than one, thus avoiding a second period of inflation.

In Fig.~\ref{fig:MN}, we report the allowed regions for $f_{\rm dec}$ and $\Omega_{\rm GW}^{\rm plt}$, once we fix the value of the leptogenesis scale $M_N$ (regions with different colors) and take the gauge coupling to vary from $10^{-7}$ to $10^{-1}$. The black lines (from left to right) represent the sensitivity of LISA, DECIGO, and ET detectors, respectively. Irrespective of the values of the gauge coupling, going from light to heavy $M_N$, the maximum allowed value for the spectral break frequency $f_{\rm dec}$ increases. Remarkably, this trend allows us to point out that a spectral break in the LISA (DECIGO) frequency band would be (most likely) a striking signature of the 3FL leptogenesis regime ($M_N \leq 10^{9}~{\rm GeV}$). On the other hand, at the ET frequency band, both the 3FL and 2FL regimes leave their signatures distinct in signal strength, with a smaller GW amplitude for the 3FL case. As shown with the BP1 (2FL) and BP2 (3FL) cases, despite corresponding to the same $f_{\rm dec}$, they have different plateau amplitudes (see Fig.~\ref{fig:GW}). On the other hand, the 2FL regime (corresponding to $g^\prime \gtrsim 10^{-3}$) is highly disfavored by the PTA data, which already excluded the 1FL case.

The GWFRL windows obtained in the present analysis are robust at the leading order, nevertheless, they can be further improved in future works. We have considered $\lambda={g^\prime}^3$, but a more rigorous parameter space scan could be performed to find a smooth transition of the field while being consistent with other gauge coupling dependent constraints (see Supplemental Material \ref{SM:gauge} for a brief discussion). Moreover, one can account for the GW emission from the higher $k$-modes. However, we find that within an instantaneous scaling network (instantaneous change in $A_\beta$ during the radiation-to-matter transition), summing a large number of modes makes the PTA constraint stronger by a factor of $\sim \zeta(4/3)$ and lowers the $f_{\rm dec}$ by a factor of $\sim 2.5$. The latter has been derived by taking a $10\%$ deviation from the near scale-invariant spectrum. In addition, the spectrum falls as $f^{-1/3}$ instead of $f^{-1}$ at very high frequencies~\cite{Blasi:2020wpy, Datta:2020bht} with no impact on our results based on the detectability of the signal at lower frequencies.

In conclusion, we have demonstrated that different flavor regimes of high-scale leptogenesis could produce GW signatures with spectral features depending on the mass acle of the heavy neutrinos.  This is an interesting aspect of the simple realization of the seesaw model based on $U(1)_{B-L}$ symmetry such that different flavored regimes could be probed without invoking any flavor symmetry to reduce the number of free parameters in the seesaw Lagrangian. On the other hand, we have established synergy between primordial gravitational waves and low-energy neutrino physics in seesaw models with flavor symmetries. It is important to emphasize that the predicted GW signal should not be regarded as a definitive indicator, as similar spectral features could arise in other BSM models involving matter domination. Furthermore, in this work, we do not delve into the high-frequency behavior of the GW spectra pertaining to different flavor regimes, nor do we perform a statistical analysis to address uncertainties in reconstructing weak GW signals (e.g., BP3 in Fig.~\ref{fig:GW}). These aspects will be thoroughly explored in an extended version of this work.

{\it \bf Acknowledgements:} We thank Tanmay Vachaspati for useful insight regarding the particle production cut-off. The work of MC, GM, RS, and NS is supported by the research project TAsP (Theoretical Astroparticle Physics) funded by the Istituto Nazionale di Fisica Nucleare (INFN). The work of NS is further supported by the research grant number 2022E2J4RK ``PANTHEON: Perspectives in Astroparticle and Neutrino THEory with Old and New messengers'' under the program PRIN 2022 funded by the Italian Ministero dell’Università e della Ricerca (MUR).  The work of SD is  supported by the National Natural Science Foundation of China (NNSFC)
under grant No. 12150610460.

\bibliography{bibliography}

\clearpage
\newpage
\maketitle
\onecolumngrid

\begin{center}
\textbf{\large Supplemental Material for \\ \documenttitle} \\ 
\vspace{0.05in}
{Marco Chianese, \ Satyabrata Datta, \ Gennaro Miele, \ Rome Samanta, and Ninetta Saviano}
\end{center}
\onecolumngrid
\setcounter{equation}{0}
\setcounter{figure}{0}
\setcounter{table}{0}
\setcounter{section}{0}
\setcounter{page}{1}
\makeatletter
\renewcommand{\theequation}{S\arabic{equation}}
\renewcommand{\thefigure}{S\arabic{figure}}
\renewcommand{\thetable}{S\arabic{figure}}

The Supplemental Material is organized as follows. In Sec.~\ref{SM:evo}, we discuss the evolution of the radiation and the scalar field energy densities, and the produced entropy with the inverse of temperature. In Sec.~\ref{SM:strings}, we justify why the cosmic strings predicted by our model can be treated as thin strings and we detail their cosmological evolution. In Sec.~\ref{SM:gauge}, we investigate the dependence of the Gravitational Waves windows for Flavored Regimes of Leptogenesis (GWFRL) on the gauge coupling.

\section{Scalar field evolution and entropy production \label{SM:evo}}

\begin{figure*}[t!]
    \centering
    \includegraphics[width=\textwidth]{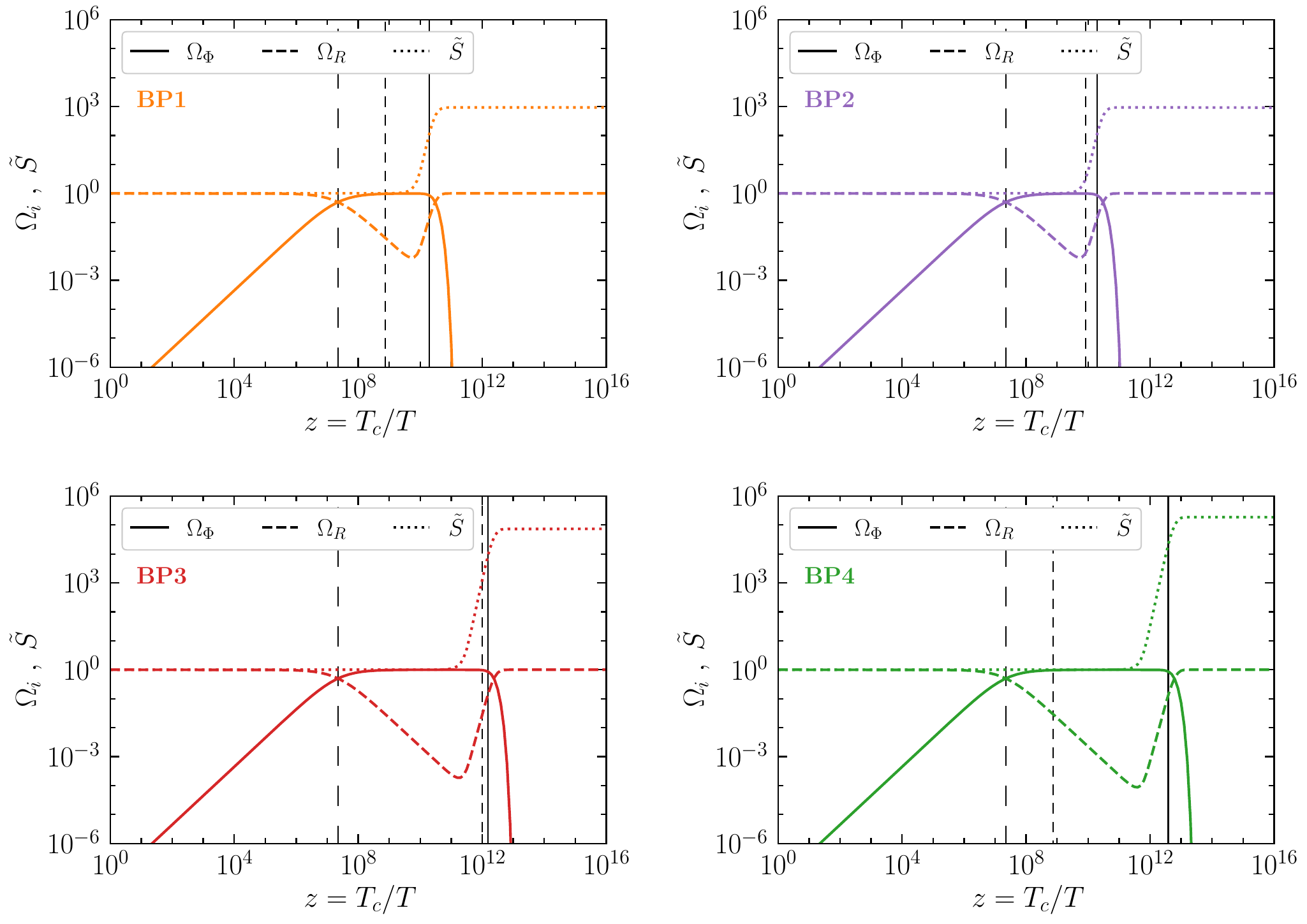}
    \caption{Evolution of the radiation and the scalar field normalized energy densities $\Omega_i=\rho_i/\rho_{\rm tot}$, and the entropy production factor $\kappa$ corresponding to the benchmarks discussed in the main text. Because only the ratio of the entropy before and after the $\Phi$ decay matters, we set $\kappa(T_c)=1$. The vertical long-dashed, dashed and solid lines represent the cut-off temperatures $T_{\rm dom}$, $T_{\rm cusp}$ and $T_{\rm dec}$, respectively. The condition $T_{\rm cusp} \geq T_{\rm dec}$ ($z_{\rm cusp} \leq z_{\rm dec}$) implies that the fall in the GW spectra occurs due to the $M_N$-dependent matter domination.}
    \label{fig:evo}
\end{figure*}

The scalar field $\Phi$ with initial vacuum energy $ \rho_{\Phi}\left(T_c \right) \simeq \lambda v_\Phi^4/4$  dominates the energy density for a period, and injects entropy as it decays. We account for this effect by solving the following Friedmann equations: 
\bea
\frac{{\rm d}\rho_R}{{\rm d}t}+4\mathcal{H}\rho_R=\Gamma_\Phi^{HH}\rho_{\Phi}\,,~~
\frac{{\rm d}\rho_{\Phi}}{{\rm d}t}+3\mathcal{H}\rho_{\Phi}=-\Gamma_\Phi^{HH}\rho_{\Phi}\,,~~
\frac{{\rm d}s}{{\rm d}t}+3\mathcal{H}s=\Gamma_\Phi^{HH}\frac{\rho_{\Phi}}{T}\,,\label{be3}
\eea
and upon recasting them as
\bea
\frac{{\rm d}\rho_{R}}{{\rm d}z}+\frac{4}{z}\rho_R=0\,, ~~
\frac{{\rm d}\rho_{\Phi}}{{\rm d}z}+\frac{3}{z}\frac{\mathcal{H}}{\tilde{\mathcal{H}}}\rho_{\Phi}+\Gamma_\Phi^{HH}\frac{1}{z\tilde{\mathcal{H}}}\rho_{\Phi}=0\,,\label{den2}
\eea
where $\rho_R$ and $\rho_\Phi$ are the energy densities of radiation and the scalar field as a function of $z=T_c/T$, $\mathcal{H}$ is the Hubble parameter, $s$ is the entropy density of the thermal bath, and the temperature-time relation has been derived from the third of Eq.~\eqref{be3} as
\bea
\frac{1}{T}\frac{{\rm d}T}{{\rm d}t}=-\left(\mathcal{H}+\frac{1}{3g_{*s}(T)}\frac{{\rm d}g_{*s}(T)}{{\rm d}t}-\Gamma_\Phi^{HH}\frac{\rho_{\Phi}}{4\rho_{R}}\right)=-\tilde{\mathcal{H}}\,,\label{temvar}
\eea
with $a$ and $g_{*s}$ being the scale factor and the number of entropy degrees of freedom, respectively. The production of entropy from the $\Phi$ decays is computed by solving 
\bea
\frac{{\rm d}a}{{\rm d}z}=\left(1+\Gamma_\Phi^{HH}\frac{\rho_{\Phi}}{4\rho_{R}\tilde{\mathcal{H}}}\right)\frac{a}{z}\,,
\eea
and computing the ratio of $\tilde{S}\sim a^3/z^3$ after and before the scalar field decay. The amount of entropy production $\kappa=\tilde{S}_{\rm after}/\tilde{S}_{\rm before}$ can be estimated as
\bea
\kappa=\frac{T_{\rm dom}}{T_{\rm dec}}\Theta\left[T_{\rm dom}-T_{\rm dec}\right]\quad{\rm with}\quad T_{\rm dom}=\frac{\rho_\Phi(T_c)}{\rho_R(T_c)}T_c\,,\label{entrp}
\eea
where $T_c$ is the critical temperature of the $U(1)_{B-L}$ phase transition, $T_{\rm dom}$ is the temperature at which the scalar field dominates the universe’s energy density, and the $\Theta$ function represents that the expression for $\kappa$ is valid for $T_{\rm dom}>T_{\rm dec}$ only. The temperature $T_{\rm dec}$ is the solution of the equation $\Gamma_\Phi^{HH}(T_{\rm dec})\simeq \mathcal{H}(T_{\rm dec})$, and takes the expression
\bea
T_{\rm dec} = \tilde{T}_{\rm dec}~\mathcal{W}\left(\frac{\Lambda}{\widetilde{T}_{\rm dec}}\right)\,, \label{newdec}
\eea
with
\bea
\widetilde{T}_{\rm dec}=\left(\frac{90}{\pi^2 g_*}\right)^{1/4} \left[\tilde{M}_{\rm Pl}~\Gamma_0~\frac{f_N^6}{\lambda}\left(\frac{v_\Phi}{\rm 10^{13}~GeV}\right)^2\left(\frac{m_\Phi}{\rm 10^8~GeV}\right)\right]^{1/2}\,,
\eea
where $\tilde{M}_{\rm Pl}=2.4\times 10^{18}~{\rm  GeV}$ is the reduced Planck constant, $g_* \simeq 106.75$ is the number of effective degrees of freedom that contribute to the radiation, and $\mathcal{W} ({\Lambda}/{\widetilde{T}_{\rm dec}})$ is a Lambert function that modulates the $T_{\rm dec}$ according to the imposed renormalisation condition.
The evolution of the radiation, the scalar field normalized energy densities $\Omega_i=\rho_i/\rho_{\rm tot}$, and the entropy production factor $\kappa$ corresponding to the benchmarks discussed in the main text are shown in Fig.~\ref{fig:evo}.

We also provide an order of magnitude estimate on the required level of quasi-degeneracy in the RH neutrinos to produce the observed baryon asymmetry. The baryon-to-photon ratio in the resonance regime of leptogenesis can be written as 
\bea
\eta_B =10^{-2}\varepsilon_i\zeta_i \kappa^{-1}\,\,\, {\rm with } \,\,\,
\varepsilon_i\simeq\frac{3M_i m_\nu}{8\pi v_h^2\delta}\,,\label{bas}
\eea
where $\varepsilon_i$ is the CP asymmetry parameter,  $\zeta_i\simeq 10^{-1}$ is the efficiency of lepton asymmetry production that takes into account the washout processes, $\delta=(M_j-M_i)/{M_i}$ is the level of quasi degeneracy in the RH neutrino masses, and $\kappa$ is the injected entropy by $\Phi$ decay as defined in Eq.~\eqref{entrp}.  Given the observed value $\eta_B\simeq 6.3\times 10^{-10}$ \cite{Planck:2015fie}, using Eq.~\eqref{bas}, we obtain 
\bea
\delta \simeq  6.25\times 10^{-1} \kappa^{-1}\left(\frac{M_N}{10^{9}\rm GeV}\right).
\eea
 Therefore, e.g., for BP4 with $\kappa\simeq 10^6$, we get $\delta \simeq 10^{-7}$ which is six orders of magnitude smaller than the standard scenarios with $\kappa=1$ (no entropy production).


\section{The string thickness and evolution\label{SM:strings}}
We name the strings produced in our model (with small $\lambda$)
as ``thick'' strings because for a given $v_\Phi$ their width $\delta_w\sim (\sqrt{\lambda} v_\Phi)^{-1}$ is larger than what is generally considered for $\lambda\simeq 1$. However, in an expanding Universe, a string is considered to be thin when $\delta_w\sim (\sqrt{\lambda} v_\Phi)^{-1}\ll \mathcal{H}(T_F)^{-1}$ is satisfied, where $\mathcal{H}(T_F)^{-1}$ is the Hubble radius at the string formation temperature $T_F$. In our model, the above condition which translates to $G\mu\ll 10^{-3} g^\prime$ with $T_F \sim T_c$, is trivially satisfied for the entire allowed parameter space. Thus, while the predicted strings in our model are technically thick, they effectively behave like thin strings. We intuitively expect that the dynamics of these predicted strings will closely resemble those typically associated with thin strings. This perspective is commonly presented in the literature addressing small $\lambda$.

Let us assume that the strings are ``thin'' so they can be treated as the commonly-considered Nambu-Goto ones. In our scenario, the scalar field does not immediately dominate the energy density, but it takes time as defined by the temperature $T_{\rm dom}$ given in Eq.~\eqref{entrp}. Therefore, the interval $T_c>T>T_{\rm dom}$ is a radiation-dominated epoch. Matter domination is obtained only within the temperature interval $T_{\rm dom}>T>T_{\rm dec}$. Typically, the string network enters the scaling regime unless the network is damped by friction defined by a temperature $T_{\rm fric}\sim \mu/M_{\rm pl}$. Hence, the network starts to emit GWs for $T_{\rm emission }\lesssim T_{\rm fric} $. In our scenario, $T_{\rm emission}$ could be either $T_{\rm emission }>T_{\rm dom}$ or $T_{\rm emission}<T_{\rm dom}$, with the preference for the former equation for majority of our parameter space. Since the evolution of the number densities of string loops from radiation to matter is well established, we can clearly evolve the network from $T>T_{\rm dom}$ to $T<T_{\rm dom}$, as the underlying process remains the same: evolving a string network from the equation of state $w=1/3$ (radiation) to $w=0$ (matter). It is crucial that $T_{\rm emission }$ remains greater than $ T_{\rm dec }$. If $T_{\rm emission}<T_{\rm dec}$, the spectral break would be dictated by $T_{\rm emission }$ instead of $ T_{\rm dec }$, with the latter deeply important for linking leptogenesis with the GW spectral features. In our scenario, we consistently have $T_{\rm emission}>T_{\rm dec}$ and the emitted GWs for $T>T_{\rm dec}$ are suppressed.

\section{Impact of the gauge coupling on the GWFRL window\label{SM:gauge}}

\begin{figure*}[h!]
    \centering
    \includegraphics[width=0.49\textwidth]{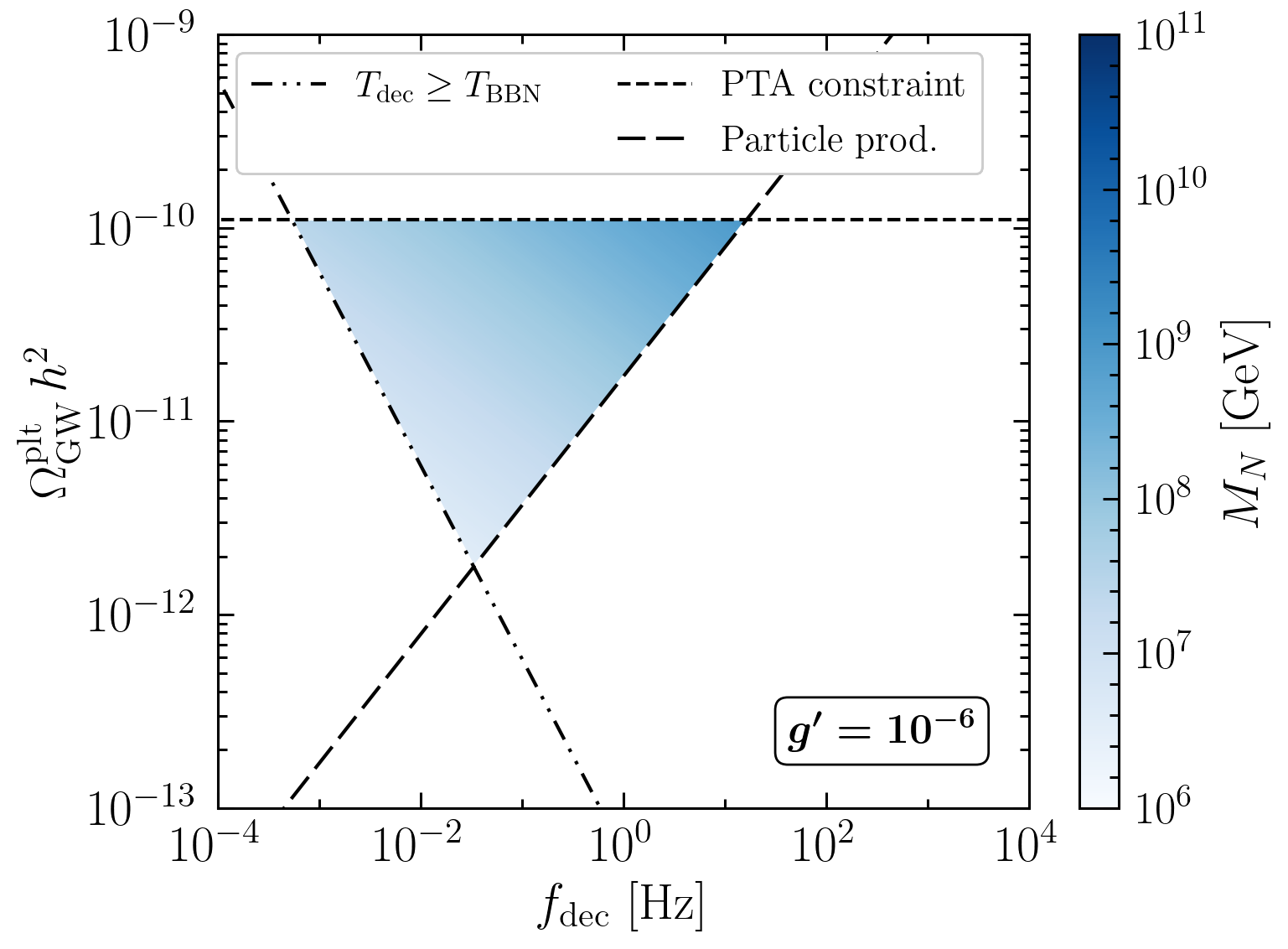}
    \includegraphics[width=0.49\textwidth]{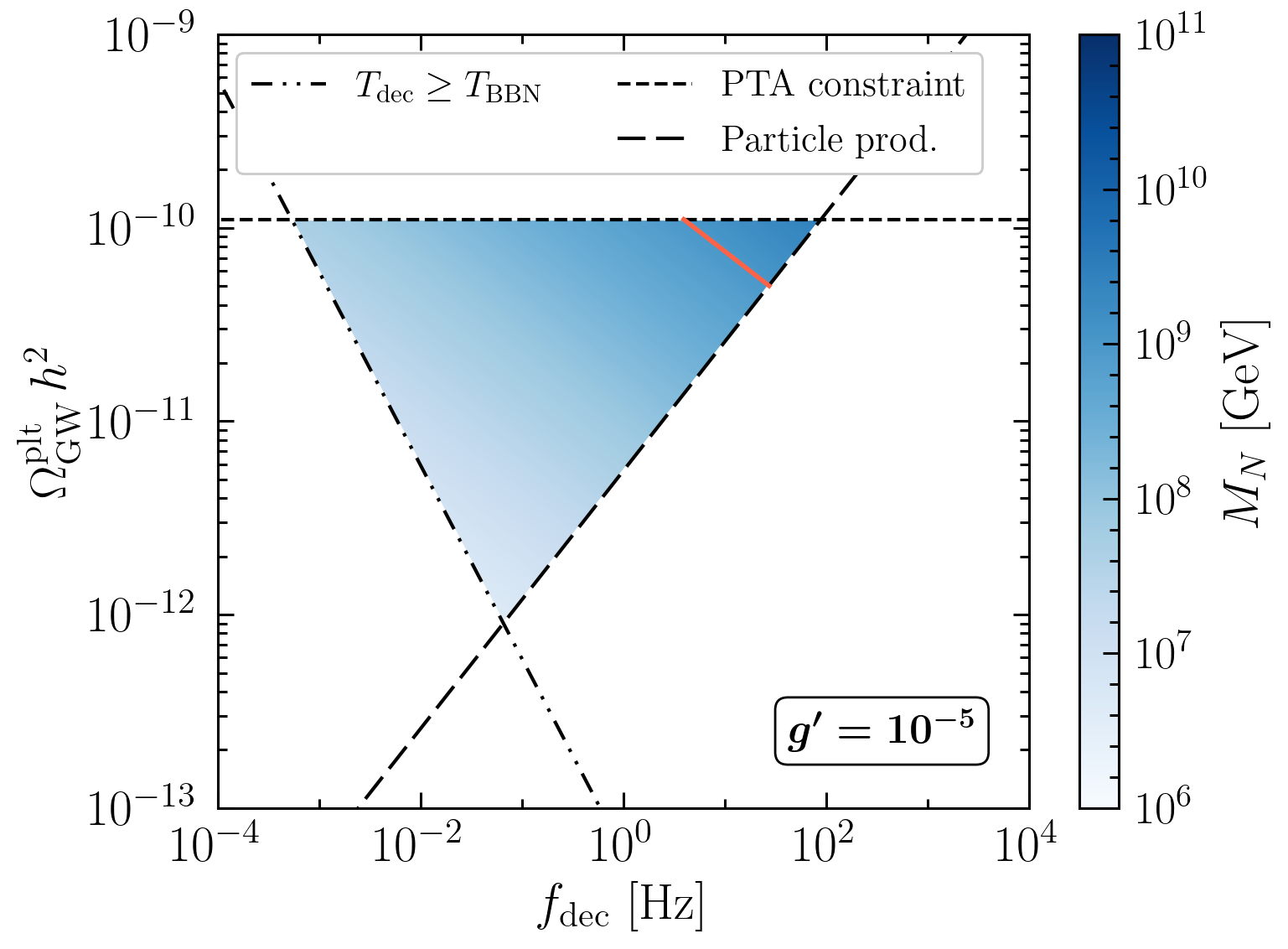}
    \includegraphics[width=0.49\textwidth]{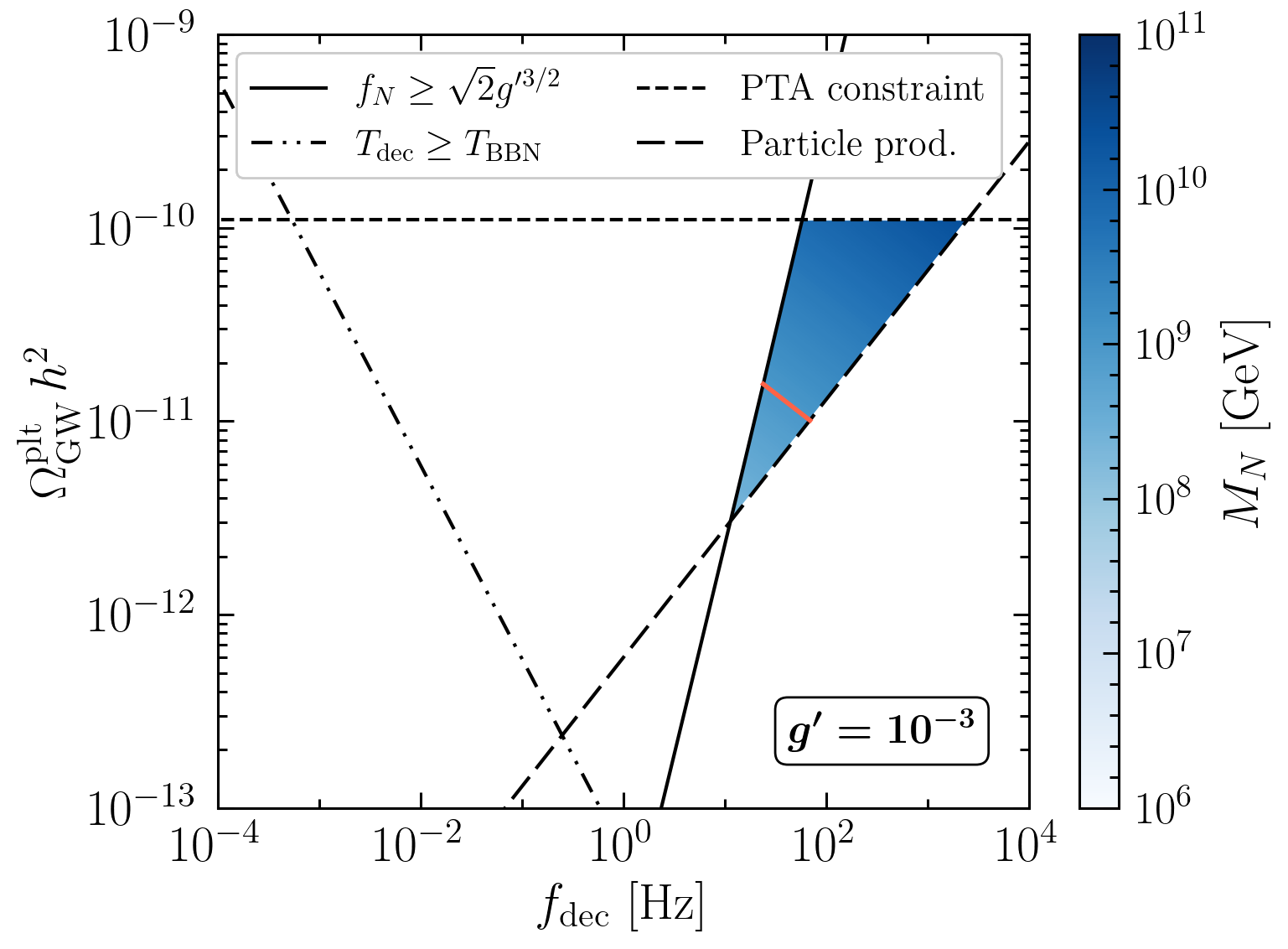}
    \includegraphics[width=0.49\textwidth]{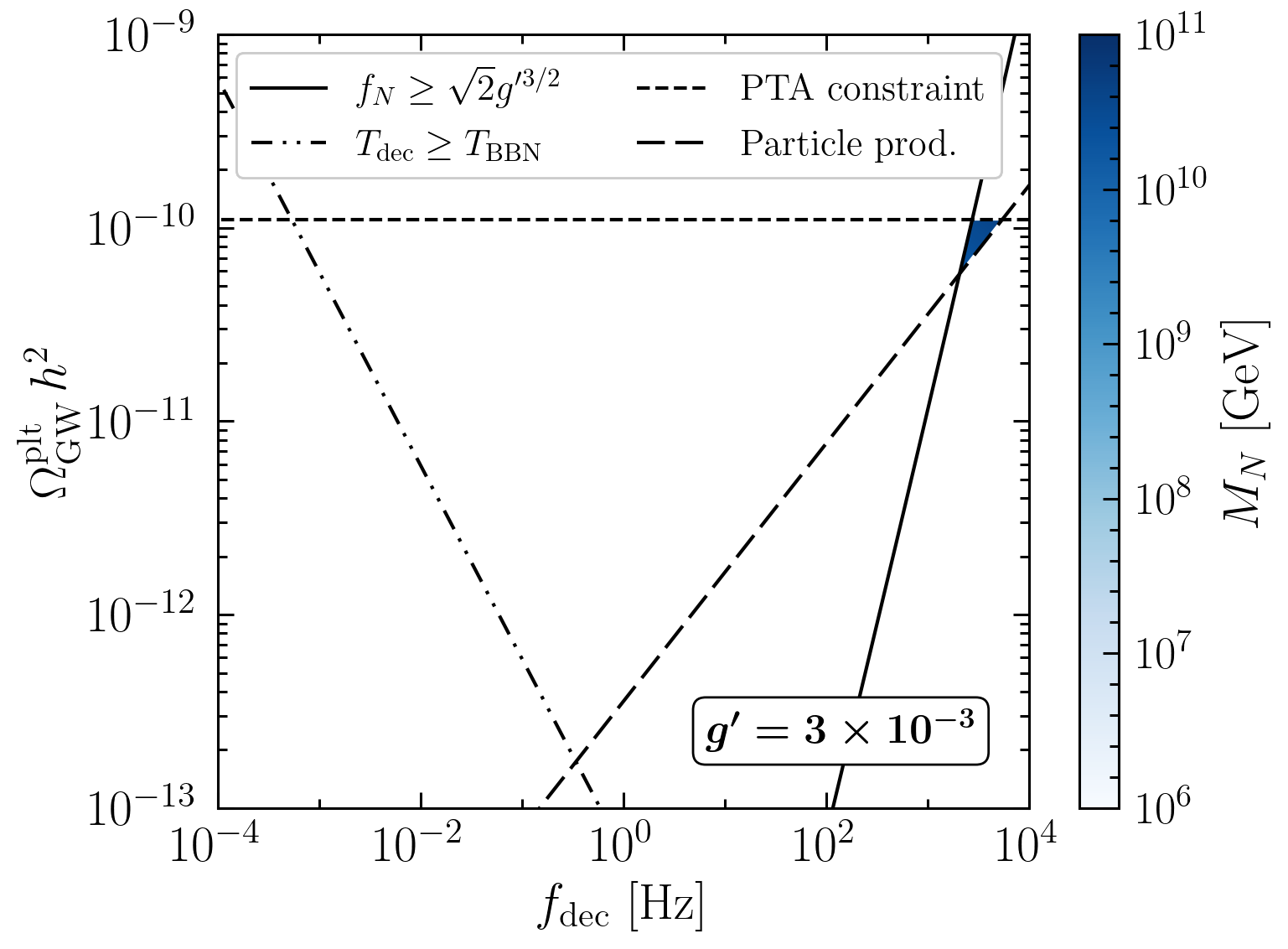}
    \caption{The allowed parameter space (GWFRL windows) in the $f_{\rm dec}$-$\Omega_{\rm GW}^{\rm plt}\,h^2$ plane for different values of the gauge coupling $g^\prime$ with the leptogenesis scale $M_N$ color-coded. The different black lines display the constraints according to the discussed model requirements and the PTA measurements. In the top panels, the constraint on the coupling $f_N$ is less restrictive being outside the displayed ranges. The orange solid line (if present) marks the transition between three-flavor and two-flavor leptogenesis regimes.}
    \label{fig:model}
\end{figure*}

In Fig.~\ref{fig:model}, we report the GWFRL windows for different choices of gauge coupling. We find that the largest allowed ranges of the spectral break frequency and the GW plateau amplitude are robust against different choices of gauge coupling. This justifies our definition of the GWFRL window in the main letter as $10^{-3} \lesssim f_{\rm dec}/{\rm Hz} \lesssim 10^3$ and $10^{-13}\lesssim\Omega_{\rm GW}^{\rm plt}h^2\lesssim 10^{-10}$. Moreover, we note that the PTA data implies that the gauge coupling is constrained to be smaller than about $3\times 10^{-3}$ (see the bottom-right panel). This further implies that the one-flavor leptogenesis regime ($M_N \geq 10^{12}~{\rm GeV}$) is already excluded by observations. A pure 3FL (2FL) regime is achieved for small (large) gauge coupling as shown in the top-left (bottom-right) panel. Both 3FL and 2FL regimes are allowed in the case of intermediate gauge coupling (top-right and bottom-left panels).

Let us also mention that the coupling hierarchy $\lambda\simeq {g^\prime}^3$ has been chosen to obtain an insignificant barrier height so that the field rolls smoothly to its minimum. A smooth transition can be only obtained for $\lambda \sim {g^\prime}^n$ with $n\lesssim 3$, while higher powers $n > 3$ would increase the barrier height. For $n < 3$, the parameter space becomes more restricted by the following constraints: {\it i)} to avoid a second period of inflation, and {\it ii)} to forbid $\Phi\rightarrow Z^\prime Z^\prime$ and $\Phi\rightarrow f\bar{f}V$ decays. Therefore, while the assumption of $\lambda \sim {g^\prime}^3$ represents the best choice of the model, we expect that all the model requirements can be satisfied with a slightly different hierarchy. Nonetheless, assuming a different hierarchy between $\lambda$ and ${g^\prime}$ would result in either obtaining a similar parameter space with different numerical values of $g^\prime$ or merely in some quantitative changes. The qualitative features in the GW spectral shape would remain unchanged because they are mostly controlled by the $f_N$ coupling.


\end{document}